\documentclass{article} 
\usepackage{hyperref}
\usepackage{url}
\usepackage{graphicx}
\usepackage[round]{natbib}
\usepackage[cmex10]{amsmath}
\usepackage{epstopdf}
\usepackage{amsmath}
\usepackage{authblk}
\DeclareMathOperator*{\argmin}{arg\,min}
\begin{document}

\title{Differential covariance: A new method to estimate functional connectivity in fMRI}

\author[1,2]{Tiger W. Lin}
\author[4]{Giri P. Krishnan}
\author[4]{Maxim Bazhenov}
\author[1,3]{Terrence J. Sejnowski}
\affil[1]{Howard Hughes Medical Institute, Computational Neurobiology Laboratory, Salk Institute for Biological Studies, La Jolla, CA 92037}
\affil[2]{Neurosciences Graduate Program, University of California San Diego, La Jolla, CA 92092}
\affil[3]{Institute for Neural Computation, University of California San Diego, La Jolla, CA 92092}
\affil[4]{Department of Medicine, University of California San Diego, La Jolla, CA 92092}
\date{}
\maketitle
\abstract

Measuring functional connectivity from fMRI is important in understanding processing in cortical networks. However, because brain's connection pattern is complex, currently used methods are prone to produce false connections. We introduce here a new method that uses derivative for estimating functional connectivity. Using simulations, we benchmarked our method with other commonly used methods. Our method achieves better results in complex network simulations. This new method provides an alternative way to estimate functional connectivity.

\section{Introduction}
\label{sec_intro}

Functional connectivity analysis is an important measurement from fMRI data \citep{friston2011functional}. It has been used quite successfully in identifying brain areas that function together in certain behaviors or at rest \citep{greicius2003functional}. Moreover, interesting geometric properties of brain networks can be studied with this method \citep{achard2006resilient}.

There are many statistical methods for estimating functional connectivity: for example, covariance-based methods, lag-based methods \citep{seth2010matlab, geweke1984measures}, Bayes-net methods \citep{meek1995causal} et al..
To compare these methods, models are used to produce simulated BOLD signals with known connection pattern. 
One popular way is to simulate neural signals from a neural network model \citep{friston2003dynamic}, then transform the signals to BOLD using an fMRI forward model (for example the nonlinear balloon model \citep{buxton1998dynamics}).
In a previous study \citep{smith2011network}, many well known statistical methods were benchmarked with a rich simulated fMRI dataset generated likewise, and covariance-based methods have been shown to be the best one in general.
  
As previously reviewed in \cite{stevenson2008inferring}, one fundamental problem for covariance-based methods is that two correlated nodes do not necessarily have direct physical connection. This is a critical problem when one wants to link the functional connectivity to the anatomical wiring of the network. 
 
We introduce a novel method to reduce these false connections and provide an estimation that is closer to the ground truth in simulated fMRI data. The new method also provides directionality information about the estimated connections. It performs better than the covariance-based methods by taking advantage of the differential signal to estimate the functional connectivity.

This paper is organized as follows: in section~\ref{sec_methods} we explain how synthetic BOLD signals were generated and introduce our new method. We also explain how results from different estimators were quantified; in section~\ref{sec_results}, we compare the performance of our new method with the performance of other methods; in section~\ref{sec_discussion}, we discuss the advantage and generalizability of our method.

\section{Methods}
\label{sec_methods}
\subsection{BOLD signal simulations}
\subsubsection{Dynamic causal modeling (DCM)}
\label{DCM}
\begin{figure}[hbtp]
\begin{center}
\includegraphics[width=\textwidth]{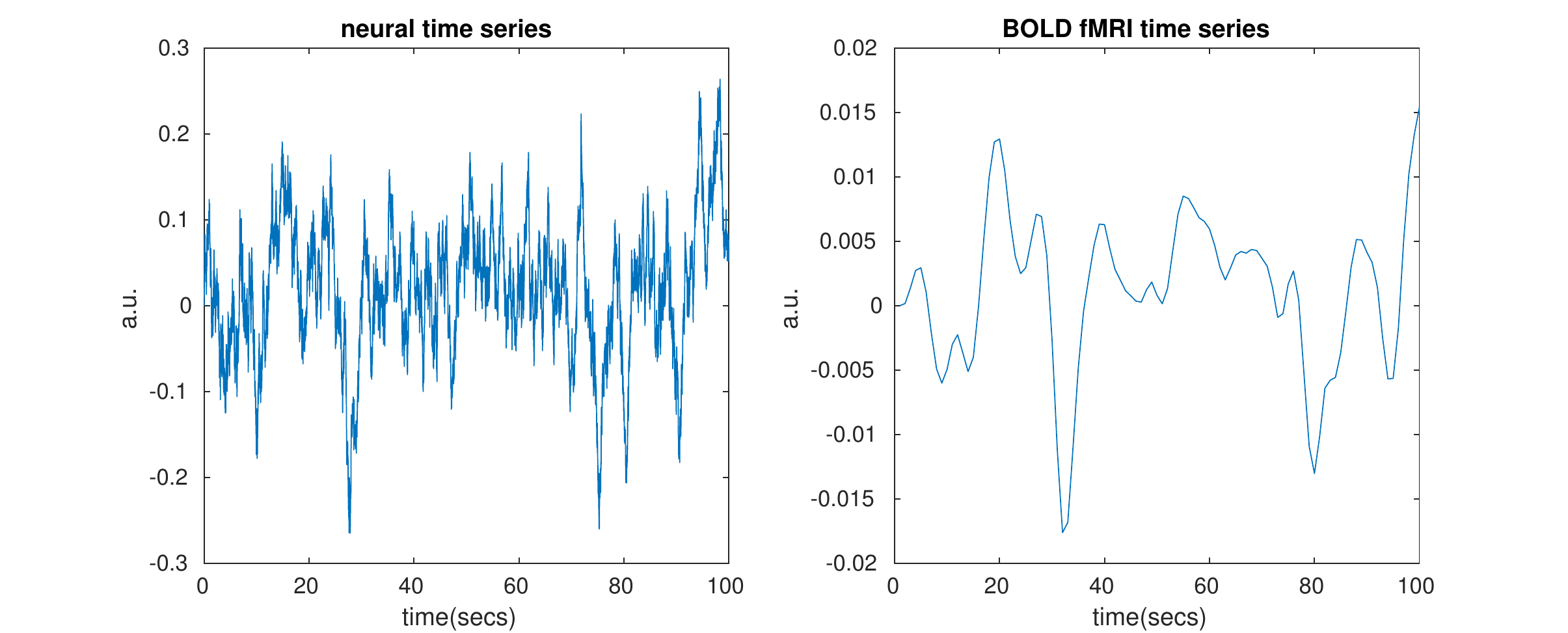}
\end{center}
\caption{Neural signal and BOLD signal example}
\label{example}
\end{figure}
We used the dynamic causal modeling \citep{friston2003dynamic} to simulate neural signals. 
The fMRI forward model is based on the Balloon model \citep{buxton1998dynamics}.
The detailed method was previously explained in \cite{smith2011network}. Briefly, the neural signals were generated from the model:
\begin{equation}
\begin{array}{lcl}
\dot{V}=AV+Cu
\end{array}
\end{equation}
where, $V$ are time series of neural signals. $A$ is the connection matrix.  Following the parameters used in the previous work \citep{smith2011network}, the diagonal values of $A$ are assigned to be -1 in our simulations.
Its non-diagonal values represent the connection between nodes and is the functional connectivity pattern we try to estimate. The connection strength is uniformly distributed between 0.4 and 0.6. An example of the connection pattern of $A$ is shown in Fig.~\ref{circuitOld}A. $A$ is a 60x60 matrix. Its first 50 nodes are visible, and they are connected with a sparse pattern. The 10 additional nodes are invisible and have a broad connection to the visible nodes. We added the latent inputs to the system because unobserved common inputs exist in real-world problems \citep{stevenson2008inferring}.
Variable $u$ is the external stimuli. For our simulations, we assume there is no external stimuli to the network. However, a white Gaussian noise with 0.1 variance is added to $u$. 
$C$ is an identity matrix in this paper indicating that each node has its independent external stimuli.

As illustrated in Fig.~\ref{example}, once the neural signals were simulated from the neural network model, they were fed into the Balloon model to generate fMRI signals. The Balloon model simulates the nonlinear vascular dynamic response due to the neural activities (\citealp{buxton1998dynamics} and  explained in \citealp{friston2003dynamic}). Here we briefly reiterate the model.
\begin{equation}
\begin{array}{lcl}
\dot{s} = V - \kappa s - \gamma(f-1)\\
\dot{f_{in}}=s\\
\tau v=f_{in} - v^{1/\alpha}\\
\tau\dot{q} = f_{in}E(f_{in},\rho)/\rho - v^{1/\alpha}q/v\\
y = V_o(k_1(1-q)+k_2(1-q/v)+k_3(1-v))
\end{array}
\label{balloon}
\end{equation}
where, $V$ is the neural signal, $s$ is the vasodilatory signal, $f_{in}$ is the blood inflow, $v$ is blood volume and $q$ is deoxyhemoglobin content. The BOLD signal $y$ is a combined signal of the blood volume and the deoxyhemoglobin content.
Lastly, the BOLD signal $y$ is sampled at 1 sec. (The value of the constants are provided in appendix~\ref{balloonApp} Table~\ref{tab_1})
\subsubsection{Benchmark dataset from previous work}
Based on the DCM, using different connection patterns, a benchmark dataset of 28 simulations was previously created \citep{smith2011network}. We used this dataset to compare our method with previous methods. To make it suitable for this work, we made several changes to the simulation based on code provided by the original authors. 

First, the original simulations used binary external stimuli with up (1) and down (0) states to simulate the ``rest'' (on average 10 secs) and the ``firing'' (on average 2.5 secs) states of the brain. In our new method, the sharp transitions between states generate large numerical derivative of the signals, and affect our method's estimation. 
The current version of our method is not capable of dealing with sharp transitions between network states, so we have to remove the binary switch of the external stimuli. 
However, a more gradual transition between two network states is acceptable by our new method. For example, the thalamocortical network model introduced below shifts among different sleep states by gradually changing the neuromodulator levels. Our method is able to give the correct estimation, even though the network model simulates different states.  

Second, we reduced the sampling rate (TR) from 3 secs to 1 sec. 
In section~\ref{backwardModel}, we developed a method  to reconstruct the original neural signals. 
From our experiments, this higher sampling rate produces better reconstruction of the neural signals, which produces better estimation results for our method. 

Third, because the new method needs more data to provide a good estimation, simulation duration was increased from 600 secs to 6000 secs for most of the simulations. 

Fourth, the thermal noise added to the BOLD signal was changed from 1\% to 0.1\%. We have to make this change because the computed derivative signals in the new method has an about 10 times lower standard deviation, thus it makes the noise tolerance of the new method about 10 times lower. All other procedures in the original paper \citep{smith2011network} were followed. 28 different connection patterns were first simulated with the DCM and passed through the balloon BOLD transfer function.  
\subsubsection{Thalamocortical model}
To further validate our method, a more realistic Hodgkin-Huxley based ionic neural network model was used in this paper to generate neural signals and the neural signals was then passed through the Balloon transfer function to generate the BOLD signals. 
The thalamocortical model used in this study was based on several
previous studies, which were used to model spindle and slow wave activity \citep{bazhenov02-8691, chen12-3987, bonjean11-9124}.
The thalamocortical model was structured as a one-dimensional, multi-layer array of cells. 
It consisted of 50 cortical pyramidal
(PY) neurons, 10 cortical inhibitory (IN) neurons, 10 thalamic relay (TC)
neurons and 10 reticular (RE) neurons.
The connection between the 50 PY neurons is shown in Fig.~\ref{HHOld}A. For the rest of the connection types, a neuron connects to all target neurons within a predefined radius as described in \citep{bazhenov02-8691}. The details of this model can be found in our previous publication \citep{lin2016diffcov}.


For each simulation, we simulated the network for 4000 secs with 600 secs at each sleep stage (Awake$\rightarrow$Stage 2$\rightarrow$Slow Wave$\rightarrow$REM$\rightarrow$Stage 2) and 200 secs of transition period between stages.
The simulated neural signals were transferred to BOLD signals and sampled at 10Hz.

\subsection{New method for functional connectivity estimation}
In our new method, we first build a backward model to reconstruct the neural signals from the BOLD signals. Then we applied our differential covariance  method to the reconstructed neural signals to estimate the functional connectivity.
\subsubsection{Backward model}
\label{backwardModel}
To derive a backward model, we first linearized the high order terms in the balloon model (eq.~\ref{balloon}) according to  \cite{khalidov2011activelets}.

Letting $\{x_1,x_2,x_3,x_4\}=\{s,1-f_{in},1-v,1-q\}$, and linearize around the resting point $\{x_1,x_2,x_3,x_4\}=\{0,0,0,0\}$, we have:
\begin{equation}
\begin{array}{lc}
\dot{x_1} = \epsilon V - x_1/\tau_s + x_2/\tau_f\\
\dot{x_2}=-x_1\\
\dot{x_3}=(x_2 - x_3/\alpha)/\tau\\
c = (1+(1-\rho)log(1-\rho)/\rho)/\tau\\
\dot{x_4}=c*x_2-(1-\alpha)/(\alpha*\tau)*x_3-x_4/\tau\\
y = V_o((k_1+k_2)*x_4+(k_3-k_2)*x_3)
\end{array}
\end{equation}


After rearrangement, we have:
\begin{equation}
\begin{array}{c}
q_1\frac{dV}{dt} + q_0V = p_4\frac{dy^4}{dt^4} + p_3\frac{dy^3}{dt^3}+ p_2\frac{dy^2}{dt^2}+ p_1\frac{dy}{dt}+p_0y  
\end{array}
\label{backwardEq}
\end{equation}
where, constants and coefficients are defined in appendix~\ref{backwardApp}.
Eq.~\ref{backwardEq} denotes that if we compute the high order derivatives of $y$, we can reconstruct this neural signal $q_1\frac{dV}{dt} + q_0V$. 

Here we denote this reconstructed signal as $z$:
\begin{equation}
\begin{array}{c}
z = q_1\frac{dV}{dt} + q_0V  
\end{array}
\end{equation}

\subsubsection{Differential covariance}
In our previous work \citep{lin2016diffcov}, we demonstrated that using derivative signals, the differential covariance  method can reduce the false connections for functional connectivity estimation. Here we briefly explain this method.

First, we assume that using the backward model above, we have reconstructed the neural signals $z(t)$ of some fMRI nodes. $z(t)$ is a NxM matrix, where N is the number of nodes recorded in the network, and M is the number of data samples during the simulation. We compute the derivative of each time series with $dz(t)=(z(t+1)-z(t-1))/(2dt)$. Then, the covariance between $z(t)$ and $dz(t)$ is computed and denoted as $\Delta_C$, which is a NxN matrix defined as the following:
\begin{equation}
\Delta_{C_{i,j}}=cov(dz_i(t),z_j(t))
\end{equation}
where $z_j(t)$ is the reconstructed neural signal of node j,  $dz_i(t)$ is the derivative neural signal of node i, and cov() is the sample  covariance function for two time series.

Note that different from our previous work \citep{lin2016diffcov}, the differential covairance is computed using the reconstructed neural signal $q_1dV/dt + q_0V$, istead of the neural signal $V$, which is not recoverable.

\subsubsection{Partial differential covariance}
\label{partial_diffCov}
As previously mentioned in \cite{stevenson2008inferring}, one problem of the covariance method is the propagation of covariance.
Here we designed a customized partial covariance algorithm to reduce this type of error in our differential covariance method.
We use $\Delta_{P_{i,j}}$ to denote the partial covariance between $dz_i(t)$ and $z_j(t)$. 

Let $Z$ be a set of all nodes except i and j.
\begin{equation}
Z=\{1,2,...,i-1,i+1,...j-1,j+1,...,N\}
\end{equation}

Using the derivation from appendix section~\ref{sec_PCOV}, we have:
\begin{equation}
\Delta_{P_{i,j}} = \Delta_{C_{i,j}} - COV_{j,Z}*COV_{Z,Z}^{-1}*\Delta_{C_{i,Z}}^T
\label{partDiffCov}
\end{equation}
where $\Delta_{C_{i,j}}$ and $\Delta_{C_{i,Z}}$ were computed from the previous section, and COV is the covariance matrix of the nodes.

\subsubsection{Sparse latent regularization}

Finally, we applied the sparse latent regularization method to our estimation \citep{chandrasekaran2011rank, yatsenko2015improved}. As explained in appendix~\ref{SLreg}, during recording, there are observable nodes and invisible nodes in a network. If the connections between the observable nodes are sparse and the number of invisible nodes is small, this method can separate the covariance into these two parts and one can regard the sparse matrix as the intrinsic connections between the observable nodes.

Here we define $\Delta_S$ as the sparse intrinsic connections from the observed nodes, and $L$ as the covariance introduced from the latent inputs. Then by solving:
\begin{equation}
\argmin\limits_{\Delta_S,L} ||\Delta_S||_1 + \alpha*tr(L)
\end{equation}
under the constraint that 
\begin{equation}
\Delta_P =\Delta_S+L
\end{equation}
we retrieve the intrinsic connections between the nodes measured.
Where, $||\ ||_1$ is the L1-norm of a matrix, and $tr()$ is the trace of a matrix. $\alpha$ is the penalty ratio between the L1-norm of $\Delta_S$ and the trace of L. It was set to 0.2 for all our estimations. $\Delta_P$ is the partial differential covariance computed from section~\ref{partial_diffCov}.

\subsection{Performance quantification}
\begin{figure}[hbtp]
\begin{center}
\includegraphics[width=\textwidth]{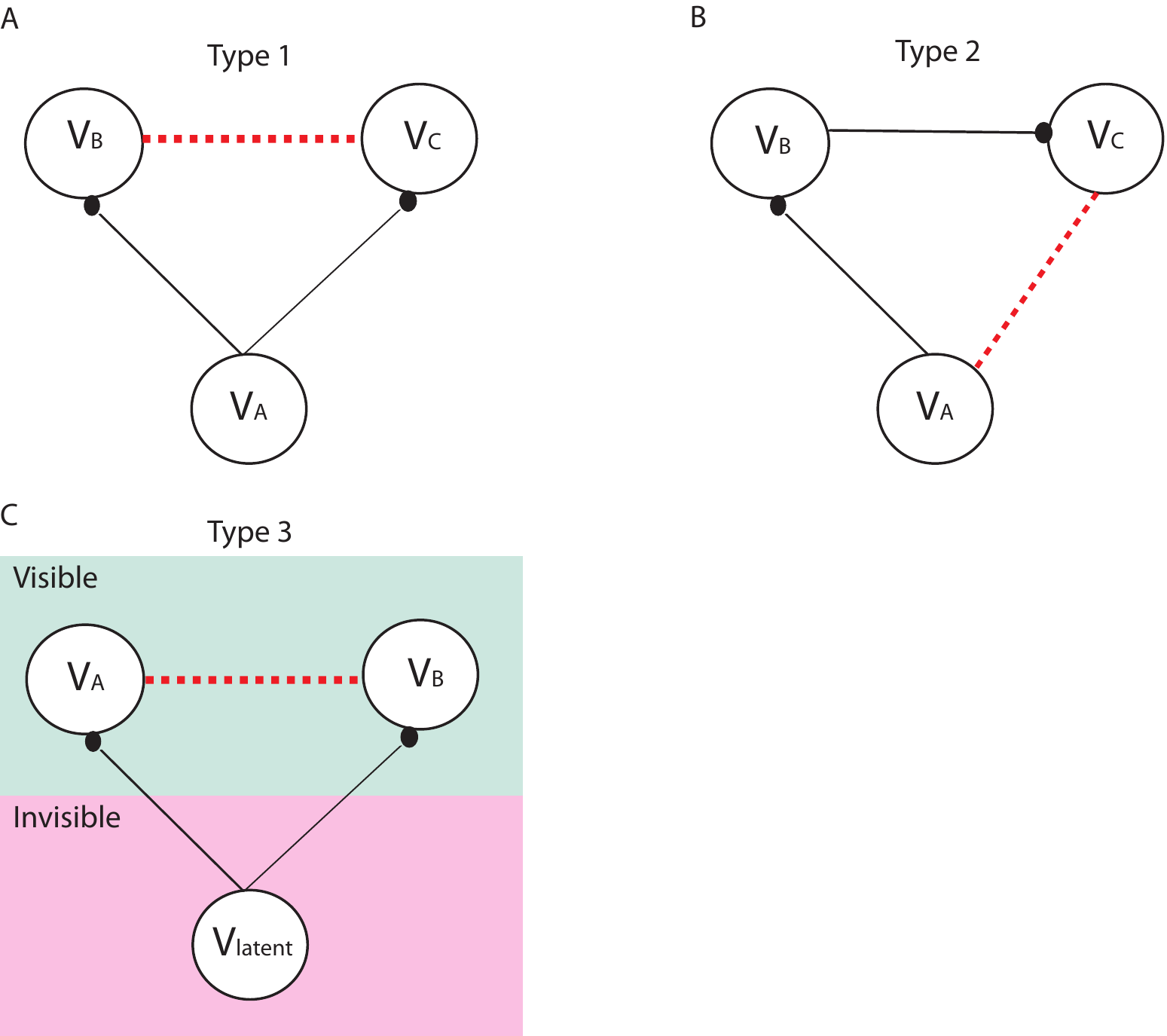}
\end{center}
\caption{Illustrations of the 3 types of false connections in the covariance-based methods. Solid lines indicate the physical wiring between neurons, and the black circles at the end indicate the synaptic contacts (i.e. the direction of the connections). The dotted lines are the false connections introduced by the covariance-based methods. A) Type 1 false connections, which are due to two neurons receiving the same synaptic inputs. B) Type 2 false connections, which are due to the propagation of correlation.  C) Type 3 false connections, which are due to unobserved common inputs.}
\label{falseConn}
\end{figure}

The performance of each method is judged by 4 quantified values. The first 3 values indicate the method's abilities to reduce the 3 types of false connections (Fig.~\ref{falseConn}). The last one indicates the method's ability to correctly estimate the true positive connections against all possible interference.

Let's define $G$ as the ground truth connectivity matrix, where:
\begin{equation}
    G_{i,j}= 
\begin{cases}
    1,  & \text{if node i projects to node j with excitatory input}\\
    -1,  &  \text{if node i projects to node j with inhibitory input}\\
    0,                 & \text{otherwise}
\end{cases}
\end{equation}

Then, we can use a 3-dimensional tensor to represent the false connections caused by common inputs. For example, node $j$ and node $k$ receive common input from node $i$:
\begin{equation}
    M_{i,j,k}= 
\begin{cases}
    1,  & \text{iff } G_{i,j}=1 \text{ and } G_{i,k}=1\\
    0,                 & \text{otherwise}
\end{cases}
\end{equation}

Therefore, we can compute a mask that labels all the type 1 false connections:
 
\begin{equation}
     mask_{1_{j,k}} = \sum_{i \in \{\text{observed nodes}\}}M_{i,j,k}
\end{equation}

For the type 2 false connections (e.g. node $i$ projects to node $k$, then node $k$ projects to node $j$), the mask is defined as:
\begin{equation}
mask_{2_{i,j}} = \sum_{k \in \{\text{observed nodes}\}}G_{i,k}G_{k,j}
\end{equation}
or, in simple matrix notation:

\begin{equation}
mask_2 = G \cdot G
\end{equation}

Similar to $mask_1$, the false connections caused by unobserved common inputs is:
\begin{equation}
    mask_{3_{j,k}} = \sum_{i \in \{\text{unobserved nodes}\}}M_{i,j,k}
\end{equation}

Lastly, $mask_4$ is defined as:

\begin{equation}
    mask_{4_{i,j}} = 
      \begin{cases}
        1,  & \text{if } G_{i,j}=0\\
        0,                 & \text{otherwise}
\end{cases}
\end{equation}

Given a connectivity matrix estimation: $Est$, and a mask: $mask_k$, ($k=1,2,3,4$),
the true positive set and the false positive set is defined as:
\begin{equation}
\begin{array}{c}
|Est_{i,j}| \in \{\text{true positive set}\}_k, \text{if } G_{i,j} \neq 0 \text{ and } mask_{k_{i,j}}=0 \\
|Est_{i,j}| \in \{\text{false positive set}\}_k, \text{if } mask_{k_{i,j}} \neq 0 \text{ and } G_{i,j}=0
\end{array}
\end{equation}

To quantify the performance, we follow the convention of \cite{smith2011network} and used the ``c-sensitivity'' index defined in the paper. Given the true positive (TP) connections and the false positive (FP) connections, c-sensitivity is defined as the percentage of the TP that is above the 95\% quantile of the FP. 

\section{Results}
\label{sec_results}
\subsection{False connections in covariance-based methods}
\begin{figure}[hbtp]
\begin{center}
\includegraphics[width=\textwidth]{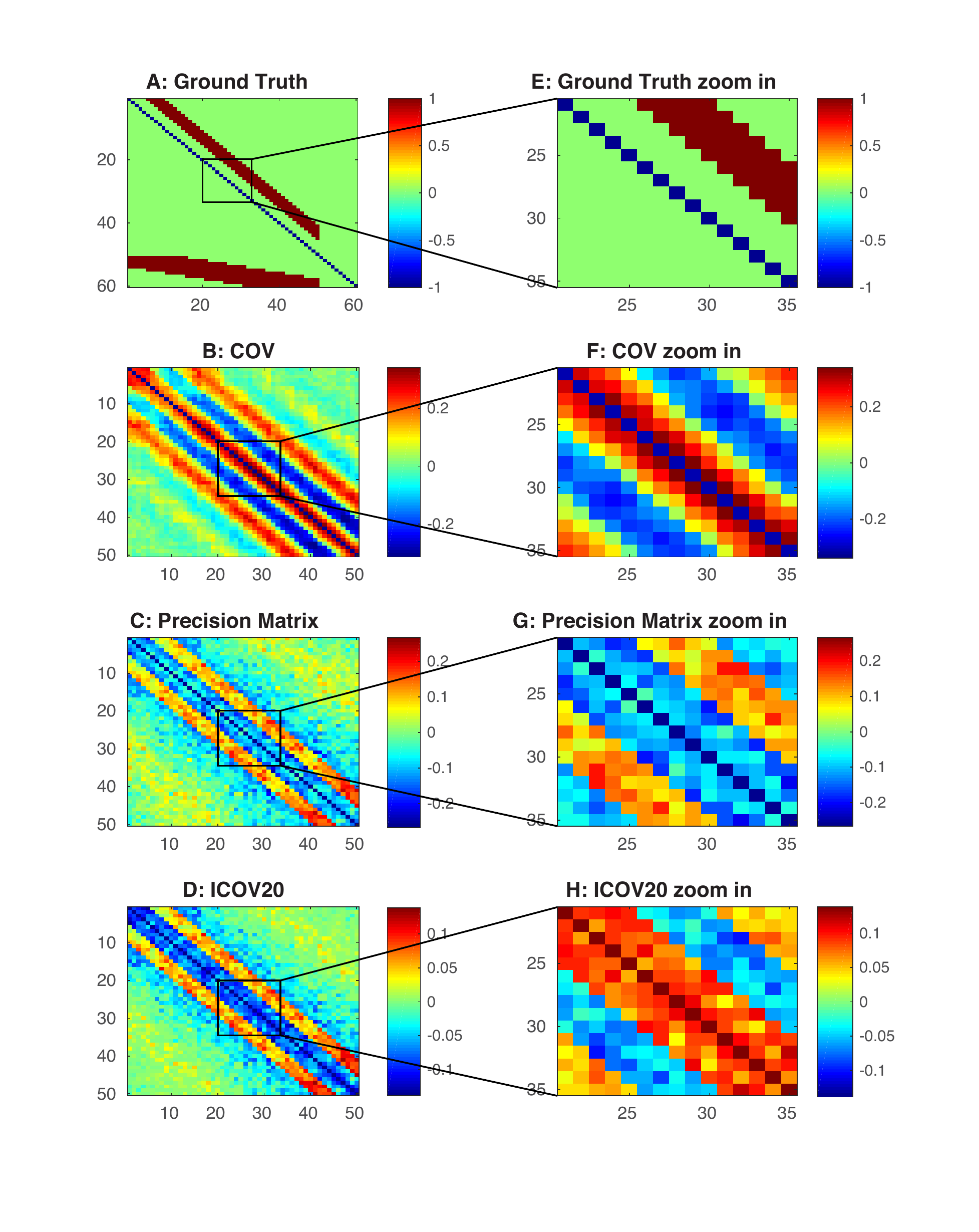}
\end{center}
\caption{DCM model's connection ground truth and connectivity estimation from covariance-based methods. A) Ground truth connection matrix. Node No.1-50 are visible nodes. Node No.51-60 are invisible nodes. B) Estimation from the covariance method. C) Estimation from the precision matrix method. D) Estimation from the sparsely regularized precision matrix (ICOV) method. E) Zoom in of panel A. F) Zoom in of panel B. G) Zoom in of panel C. H) Zoom in of panel D.}
\label{circuitOld}
\end{figure}
The commonly used covariance-based methods produce systematic false connections, and this is a well known problem \citep{stevenson2008inferring}. 
Shown in Fig.~\ref{circuitOld}A is the ground truth of the connection in our DCM model (Node No.1-50 are the observable nodes). Fig.~\ref{circuitOld}B is from the covariance method, Fig.~\ref{circuitOld}C is the precision matrix, and Fig.~\ref{circuitOld}D is the ICOV (sparsely regularized precision matrix). Comparing with the ground truth, all of these methods produce extra false connections. Here we briefly review the cause of these 3 types of false connections. The detailed mechanisms are well reviewed before \citep{stevenson2008inferring}.

For the convenience of explanation, we define the diagonal strip of connections in the ground truth (Fig.~\ref{circuitOld}A) as the +6 to +10 diagonal lines, because they are 6 to 10 steps away from the diagonal line of the matrix.

Shown in Fig.~\ref{falseConn}A, the first type of false connections are produced because two nodes receive the same input from another node. 
The same input that passes into the two nodes generates positive correlation between the two nodes. 
However, there is no physiological connection between these two nodes.
For example, as shown in  Fig.~\ref{circuitOld}A, node 1 projects to node 6 and node 7 ($A_{16}, A_{17}>0$), but node 6 and node 7 are not connected. However, because they share the same input from node 1, in Fig.~\ref{circuitOld}B, there are positive connections on coordinate (6,7) and (7,6). Because this connection pattern is also applied to other nodes (node 2 projects to node 7 and node 8, node 3 projects to node 8 and so forth)
There are false connections on the $\pm1$ to $\pm4$ diagonal lines of Fig.~\ref{circuitOld}B.

Shown in Fig.~\ref{falseConn}B, the second type of false connections are due to the propagation of the covariance. 
Because node $A$ connects to node $B$ and node $B$ connects to node $C$,  the covariance method presents covariance between node $A$ and node $C$, which do not have a physical connection. This phenomenon is shown in Fig.~\ref{circuitOld}B as the extra diagonal strips.

Shown in Fig.~\ref{falseConn}C, the third type of false connections are also due to the common input passed into two nodes. 
However, in this case, they are from the unobserved latent inputs. For this DCM model, it is due to the inputs from the 10 invisible nodes (node No.51-60) as shown on Fig.~\ref{circuitOld}A.
Because the latent nodes have broad connections to the observed nodes, they introduce a square box shape covariance pattern into  Fig.~\ref{circuitOld}B. 
And this type of false connections remains in the precision matrix and ICOV estimations (Fig.~\ref{circuitOld}C, D) as the yellow color false connections on the bottom-left corner and the up-right corner.

\subsection{Estimation from differential covariance method}
\begin{figure}[hbtp]
\begin{center}
\includegraphics[width=\textwidth]{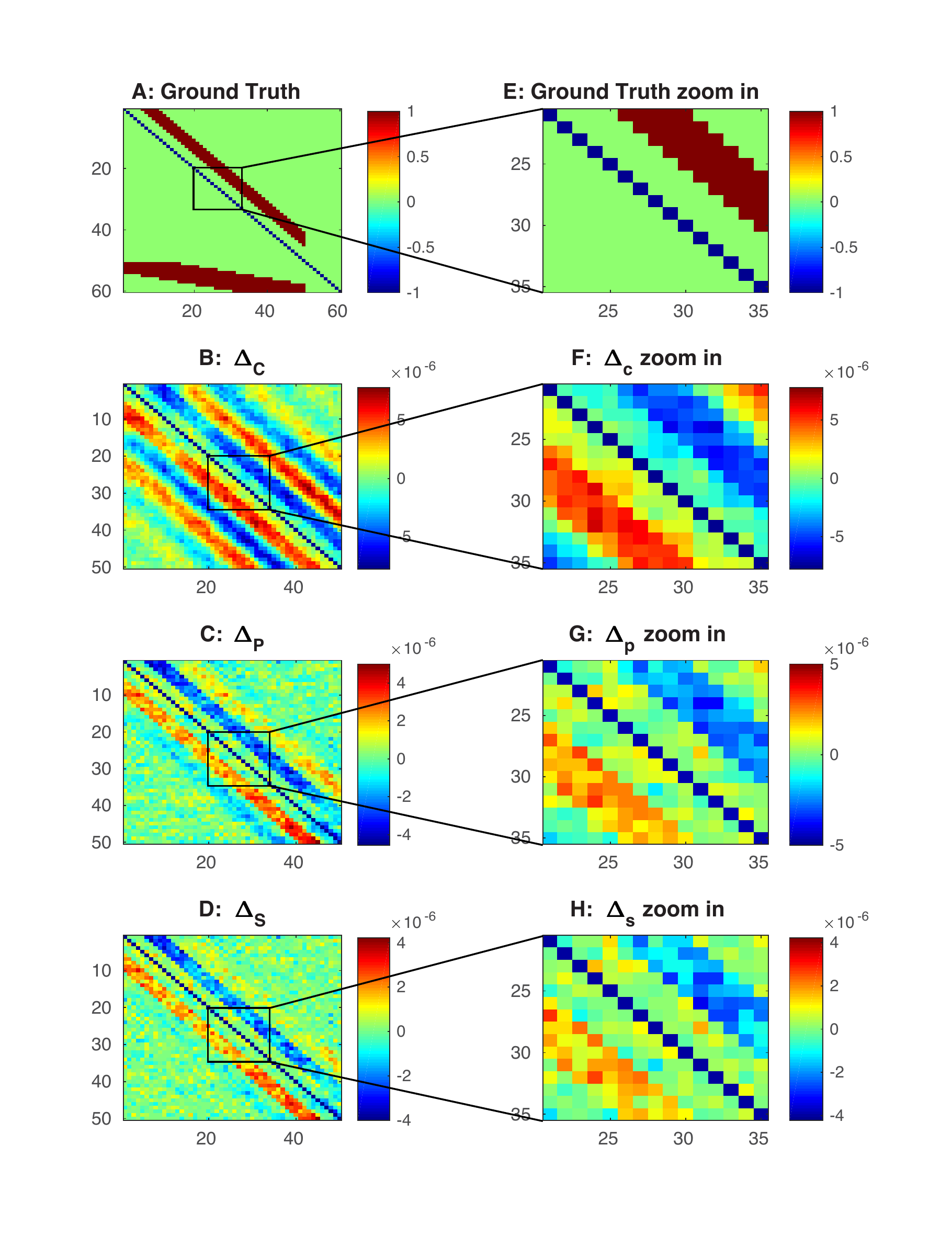}
\end{center}
\caption{Differential covariance analysis of the DCM model. The color in B,C,D,F,G,H indicates direction of the connections. For element $A_{ij}$, warm color indicates $i$ is the sink, $j$ is the source, i.e. $i \leftarrow j$, and cool color indicates $j$ is the sink, $i$ is the source,  i.e. $i \rightarrow j$. A) Ground truth connection matrix.  B) Estimation from the differential covariance method. C) Estimation from the partial differential covariance. D) Estimation from the sparse+latent regularized partial differential covariance. E) Zoom in of panel A. F) Zoom in of panel B. G) Zoom in of panel C. H) Zoom in of panel D.}
\label{circuitNew}
\end{figure}

\begin{figure}[hbtp]
\begin{center}
\includegraphics[width=\textwidth]{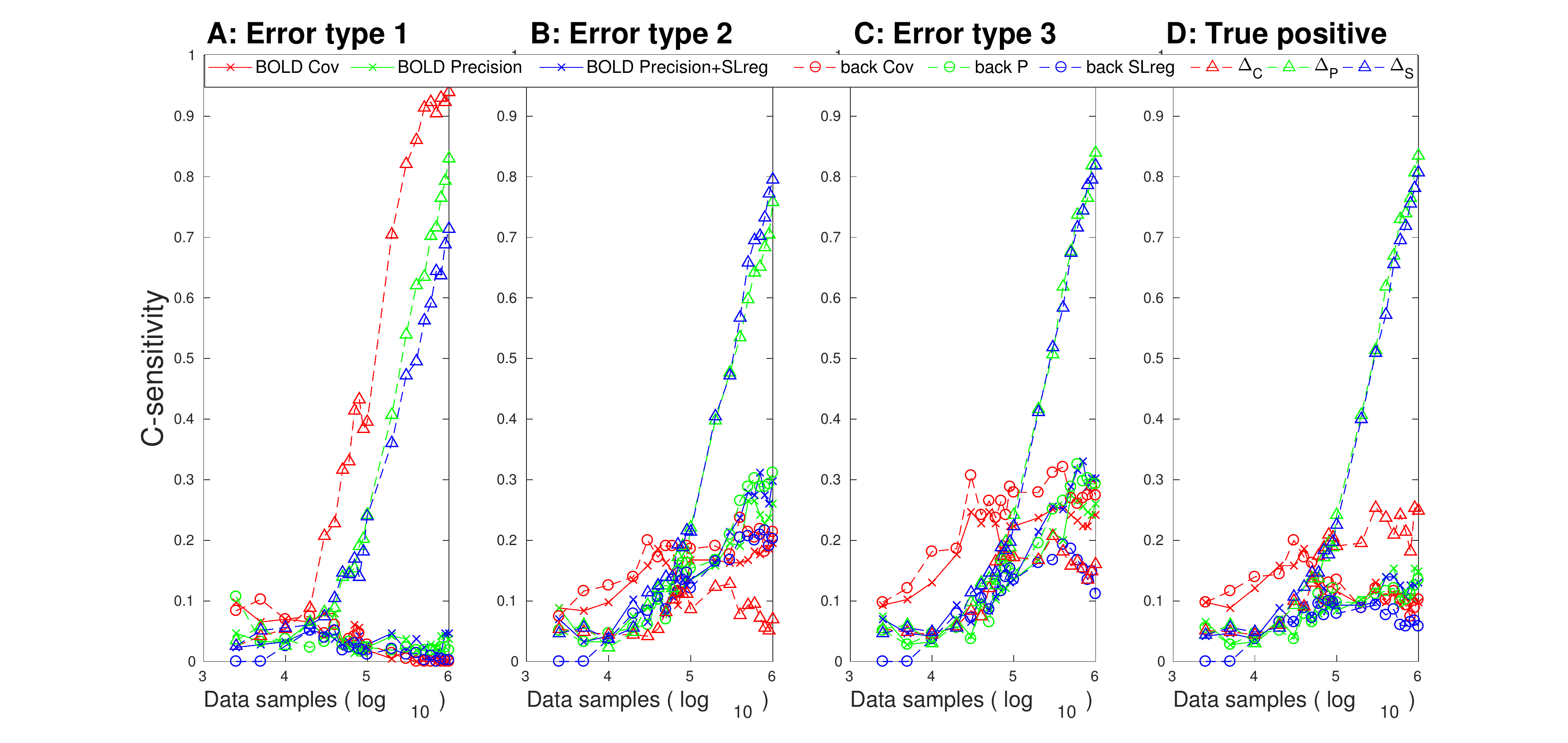}
\end{center}
\caption{Performance quantification (c-sensitivity) of different methods with respect to their abilities to reduce the 3 types of false connections and their abilities to estimate the true positive connections using the dynamic causal modeling. The first three results are from covariance-based methods directly applied to simulated BOLD signals. The next three results are from covariance-based methods applied to backward model reconstructed neural signals. The last three results are from differential covariance methods applied to backward model reconstructed neural signals. }
\label{circuit_benchmark}
\end{figure}

Below we explain the estimation results from the differential covariance  method.
Comparing the ground truth connections in Fig.~\ref{circuitNew}A with our final estimation in Fig.~\ref{circuitNew}D, we see that our method essentially transformed the connections in the ground truth into a map of sources and sinks in a network. An excitatory connection, $i \rightarrow j$, in our estimations have negative value for $\Delta_{S_{ij}}$ and positive value for $\Delta_{S_{ji}}$, which means the current is coming out of the source $i$, and goes into the sink $j$. We note that there is another ambiguous case, an inhibitory connection $j \rightarrow i$, which produces the same results in our estimations. 
Our method can not differentiate these two cases, instead, they indicate sources and sinks in a network.

\subsubsection{The differential covariance method reduces type 1 false connections}
Comparing Fig.~\ref{circuitOld}B and Fig.~\ref{circuitNew}B, we see that the type 1 false connections on $\pm1$ to $\pm4$ diagonal lines of Fig.~\ref{circuitOld}B are reduced in Fig.~\ref{circuitNew}B. This is quantified in Fig.~\ref{circuit_benchmark}A, where the differential covariance  method has higher c-sensitivity score given the type 1 false connections as the false positive set.

\subsubsection{The partial differential covariance reduces type 2 false connections}
Second, we see in Fig.~\ref{circuitNew}B, there are propagation errors. By applying the partial covariance method, we regress out the interfering terms.
Thus, there is less propagation of covariance (type 2 false connections mentioned above) in Fig.~\ref{circuitNew}C. Each method's performance for reducing type 2 false connections is quantified in Fig.~\ref{circuit_benchmark}B.

\subsubsection{The sparse+latent regularization reduces type 3 false connections}
Third, we use the sparse latent regularization to remove the covariance introduced by the latent variables. As mentioned in the method section, when the observable nodes' connections are sparse and the number of latent inputs is small, covariance introduced by the latent inputs can be separated. 
As shown in Fig.~\ref{circuitNew}D, the external covariance in the background of Fig.~\ref{circuitNew}C is reduced, while the true positive connections and the directionality of the connection is maintained. Each method's ability to reduce this type of error is quantified in Fig.~\ref{circuit_benchmark}C, and each method's overall performance at reducing all errors is quantified in Fig.~\ref{circuit_benchmark}D.

%

\subsection{Thalamocortical model's results}
\begin{figure}[hbtp]
\begin{center}
\includegraphics[width=\textwidth]{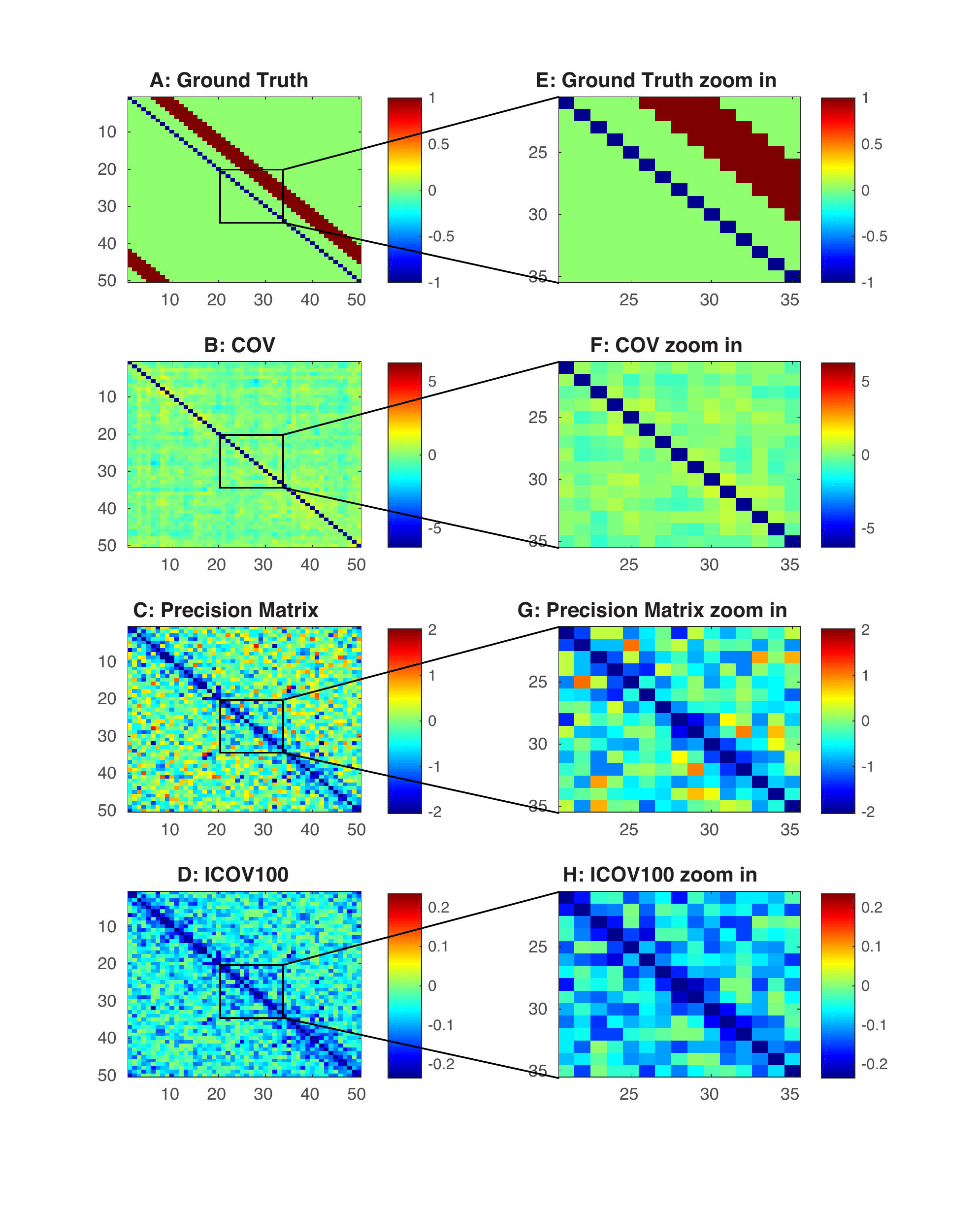}
\end{center}
\caption{BOLD signals from the thalamocortical model with connectivity estimations from the covariance-based methods. A) Ground truth connection of the PY neurons in the thalamocortical model. B) Estimation from the covariance method. C) Estimation from the precision matrix method. D) Estimation from the sparsely regularized precision matrix (ICOV) method. E) Zoom in of panel A. F) Zoom in of panel B. G) Zoom in of panel C. H) Zoom in of panel D.}
\label{HHOld}
\end{figure}
\begin{figure}[hbtp]
\begin{center}
\includegraphics[width=\textwidth]{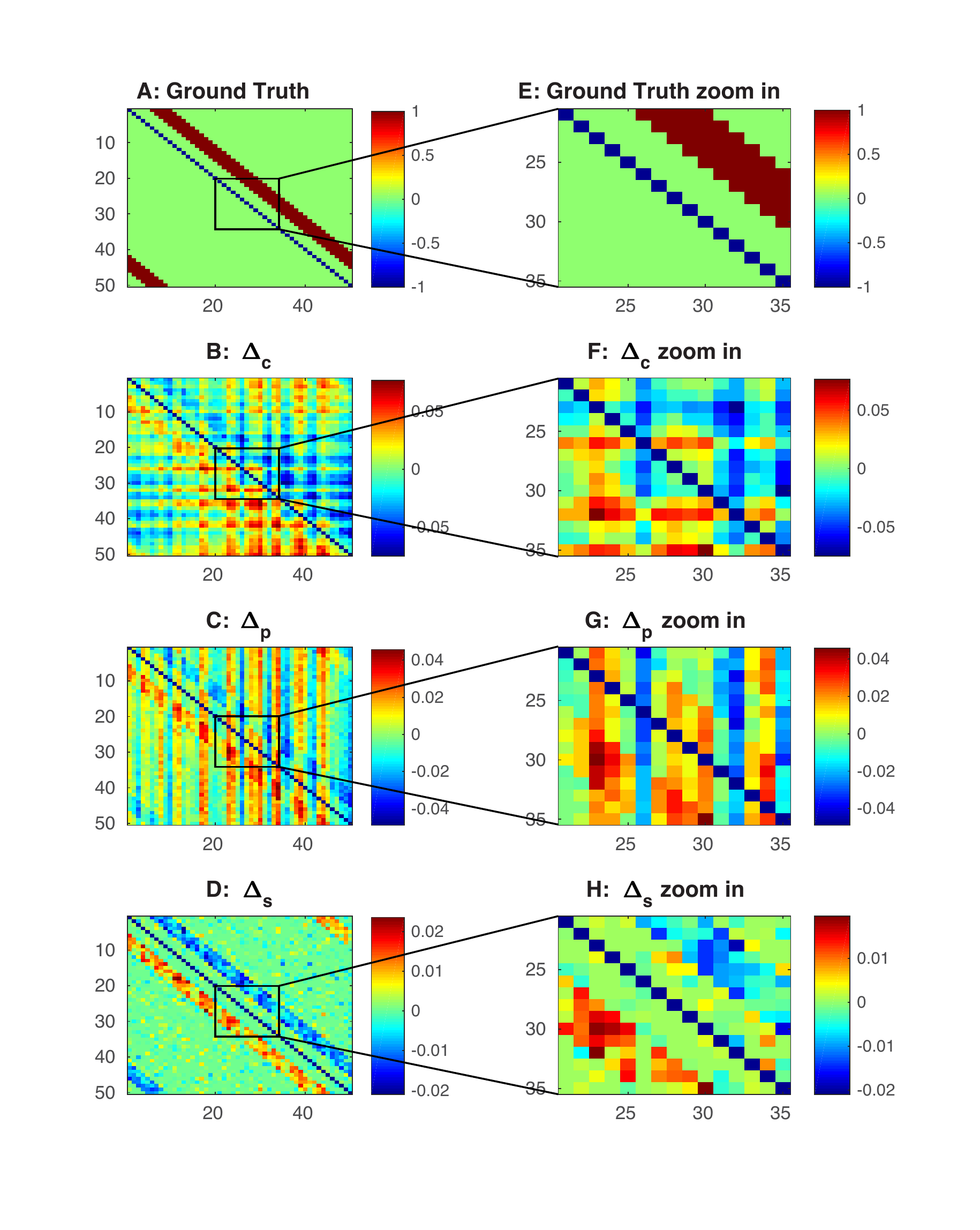}
\end{center}
\caption{Differential covariance analysis of the thalamocortical model's reconstructed neural signals. The color in B,C,D,F,G,H indicates direction of the connections. For element $A_{ij}$, warm color indicates $i$ is the sink, $j$ is the source, i.e. $i \leftarrow j$, and cool color indicates $j$ is the sink, $i$ is the source,  i.e. $i \rightarrow j$. A) Ground truth connection matrix. B) Estimation from the differential covariance method. C) Estimation from the partial differential covariance method. D) Estimation from the sparse+latent regularized partial differential covariance method. E) Zoom in of panel A. F) Zoom in of panel B. G) Zoom in of panel C. H) Zoom in of panel D.}
\label{HHNew}
\end{figure}
In addition to the DCM simulations, we also benchmarked various methods with a more realistic Hodgkin-Huxley based model.
The  covariance-based methods are  applied to both the raw BOLD signals and the reconstructed neural signals. The differential covariance method is applied to both the raw BOLD signals and the reconstructed neural signals as well.
Compared to the ground truth in Fig.~\ref{HHOld}A, covariance-based  estimations failed to show the true connections. Also the type 1 false connections are noticeable on the $\pm1$ diagonal line of Fig.~\ref{HHOld} C, and D.
We show later that, applying the covariance-based methods to the reconstructed neural signals does provide better results, however, covariance-based methods still suffer from the 3 types of false connections mentioned above.

\begin{figure}[hbtp]
\begin{center}
\includegraphics[width=\textwidth]{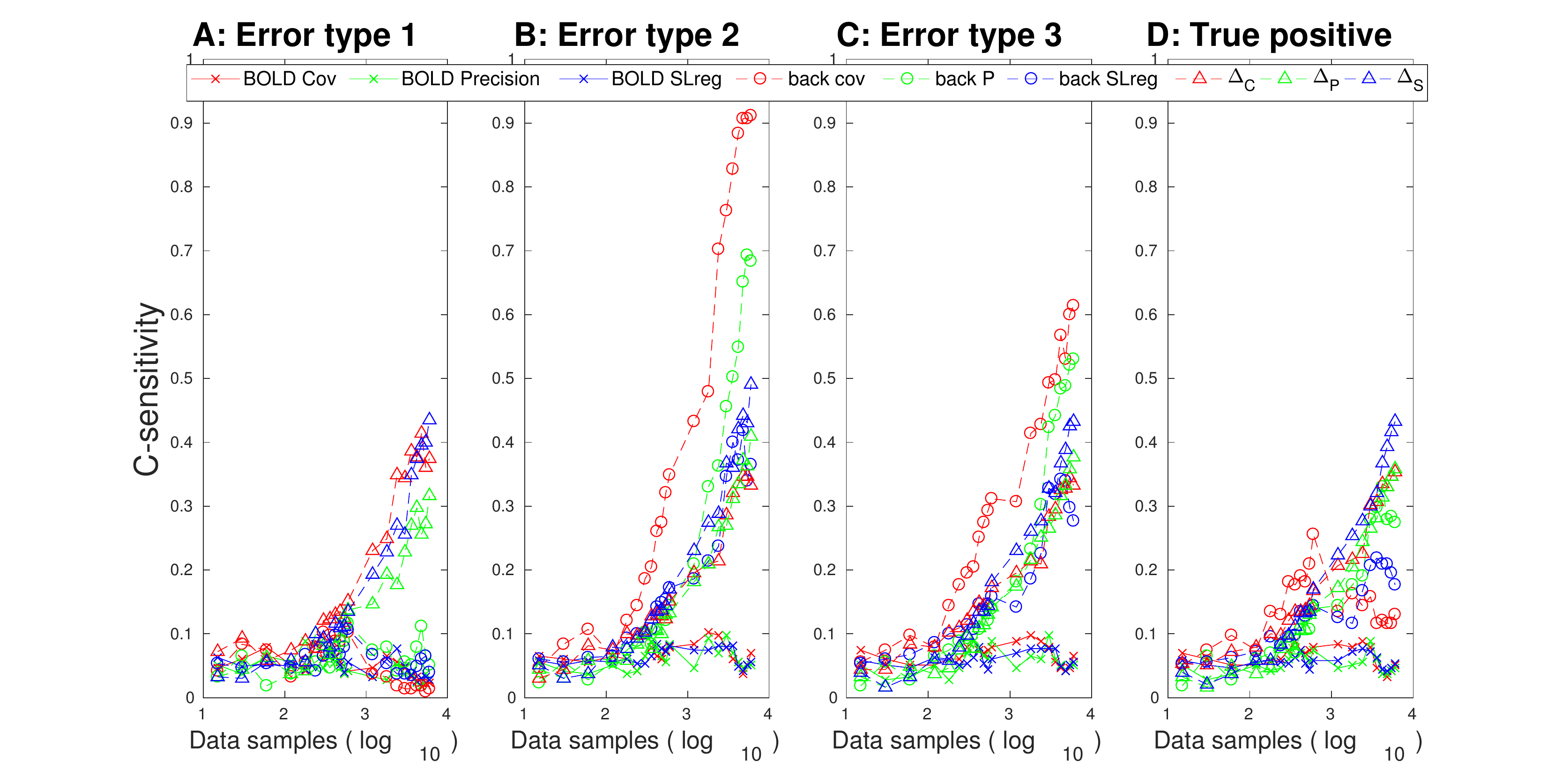}
\end{center}
\caption{Performance quantification (c-sensitivity) of different methods with respect to their abilities to reduce the 3 types of false connections and their abilities to estimate the true positive connections using the thalamocortical model.  The first three results are from covariance-based methods directly applied to simulated BOLD signals. The next three results are from covariance-based methods applied to backward model reconstructed neural signals. The last three results are from differential covariance methods applied to backward model reconstructed neural signals.}
\label{HH_benchmark}
\end{figure}

On the other hand, by applying differential covariance to the reconstructed neural signals, we see in Fig.~\ref{HHNew}B, $\Delta_C$ provides directionality information of the connection. However, $\Delta_C$ estimation is contaminated by connections introduced from the propagation of the signals and the latent inputs. The false connections due to the common current (type 1 false connections) are greatly reduced. In Fig.~\ref{HHNew}C, the partial covariance reduces the type 2 false connections and produces $\Delta_P$. Finally, in Fig.~\ref{HHNew}D, the sparse+latent regularization is applied to reduce the false connections from the latent inputs.

In Fig.~\ref{HH_benchmark}, each estimator's performance on each type of false connections is quantified. As explained above, while covariance-based methods perform well at reducing type 2 and type 3 false connections, differential covariance method  performs better at reducing type 1 false connections. Thus it achieves better performance overall.

\subsection{The backward model and the differential covariance are both needed to produce correct estimations}

\begin{figure}[hbtp]
\begin{center}
\includegraphics[width=\textwidth]{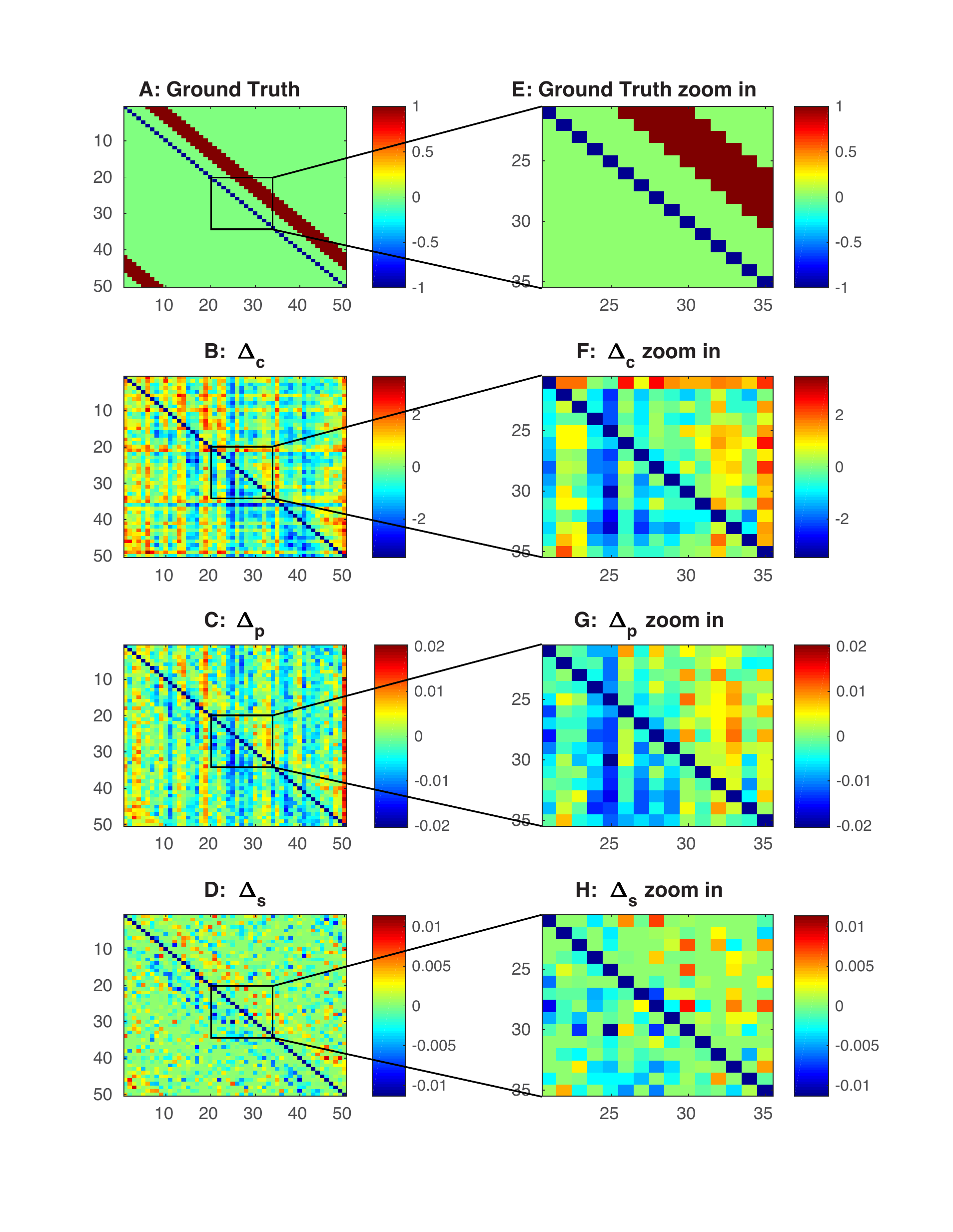}
\end{center}
\caption{Differential covariance method applied directly to simulated BOLD signals without applying the backward model. The color in B,C,D,F,G,H indicates direction of the connections. For element $A_{ij}$, warm color indicates $i$ is the sink, $j$ is the source, i.e. $i \leftarrow j$, and cool color indicates $j$ is the sink, $i$ is the source,  i.e. $i \rightarrow j$. A) Ground truth connection matrix. B) Estimation from the differential covariance method. C) Estimation from the partial differential covariance method. D) Estimation from the sparse+latent regularized partial differential covariance method. E) Zoom in of panel A. F) Zoom in of panel B. G) Zoom in of panel C. H) Zoom in of panel D.}
\label{control1}
\end{figure}

Above, we explained the function of our newly developed backward model and the differential covariance method. As a control experiment, we show here that both steps are necessary to produce good estimations. 

The purpose of the backward model is to restore the dynamic forms of the neural signal, such that it is suitable for the differential covariance estimation. As shown in Fig.~\ref{control1}, if we directly apply the differential covariance method to the raw BOLD signals, the estimator's performance is poor.

\begin{figure}[hbtp]
\begin{center}
\includegraphics[width=\textwidth]{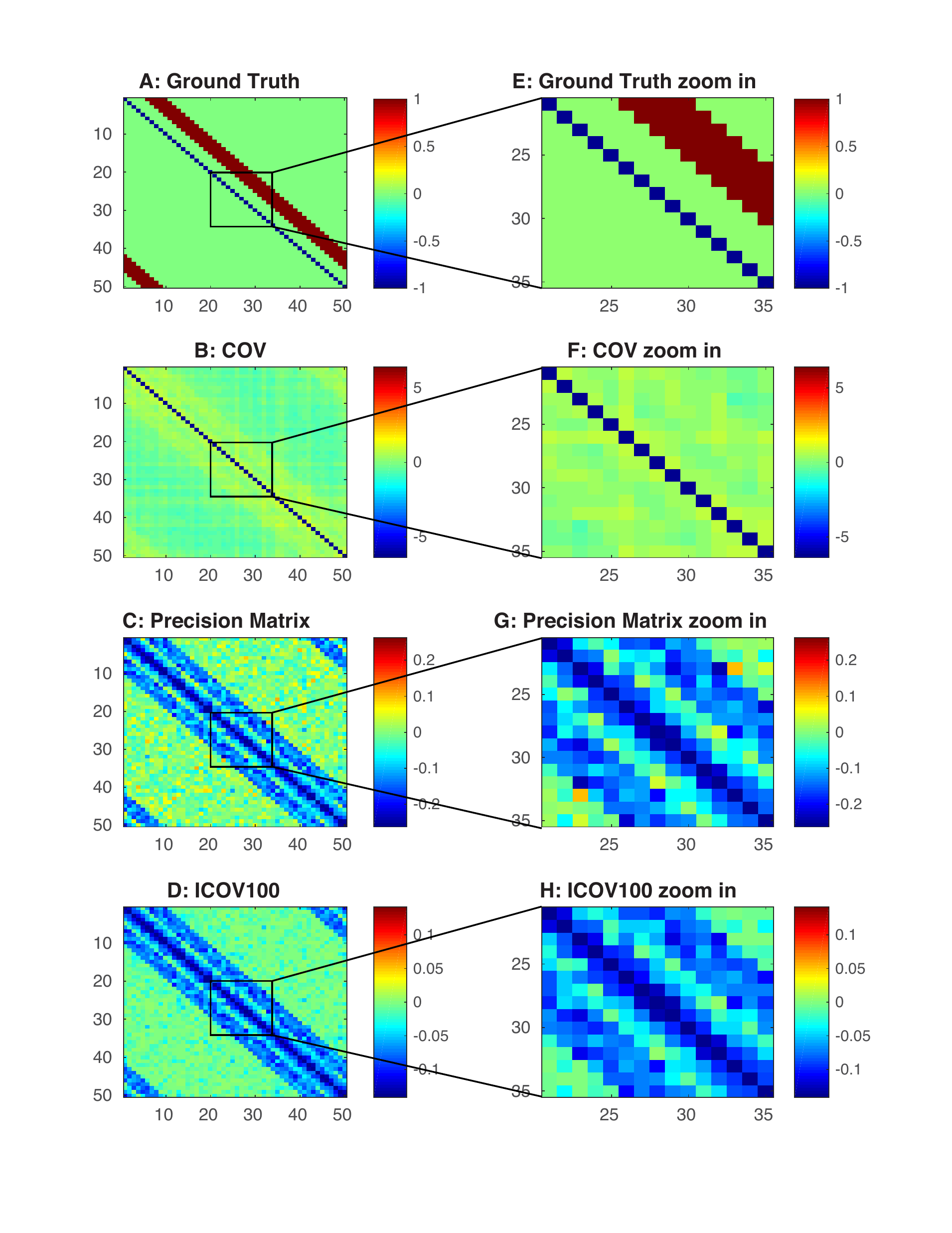}
\end{center}
\caption{covariance-based methods applied to backward model reconstructed signals. A) Ground truth connection matrix. B) Estimation from the covariance method. C) Estimation from the precision matrix method. D) Estimation from the sparsely regularized precision matrix (ICOV) method. E) Zoom in of panel C. F) Zoom in of panel D. E) Zoom in of panel A. F) Zoom in of panel B. G) Zoom in of panel C. H) Zoom in of panel D.}
\label{control2}
\end{figure}

On the other hand, the differential covariance method is also needed to reduce the false connections. In Fig.~\ref{control2}, estimations from the covariance-based methods applied to the reconstructed neural signals failed to reduce the false connections mentioned above. 
Particularly, the 1st type false connections that are on the $\pm1$ and $\pm2$ diagonal lines are apparent in Fig.~\ref{control2}C and D. 
This comparison can also be seen from the quantifications on Fig.\ref{circuit_benchmark} and Fig.\ref{HH_benchmark}. For both simulations, even using the same backward processed signals, the differential covariance method performs better at reducing the type 1 false connections than the covariance-based methods. While the covariance-based methods perform well at reducing the type 2 and type 3 false connections, differential covariance  method's performance is better overall.
Therefore, the differential covariance method is needed to make better estimations after the backward processing step.

\subsection{Benchmark using previous study's dataset}

\begin{figure}[hbtp]
\begin{center}
\includegraphics[width=\textwidth]{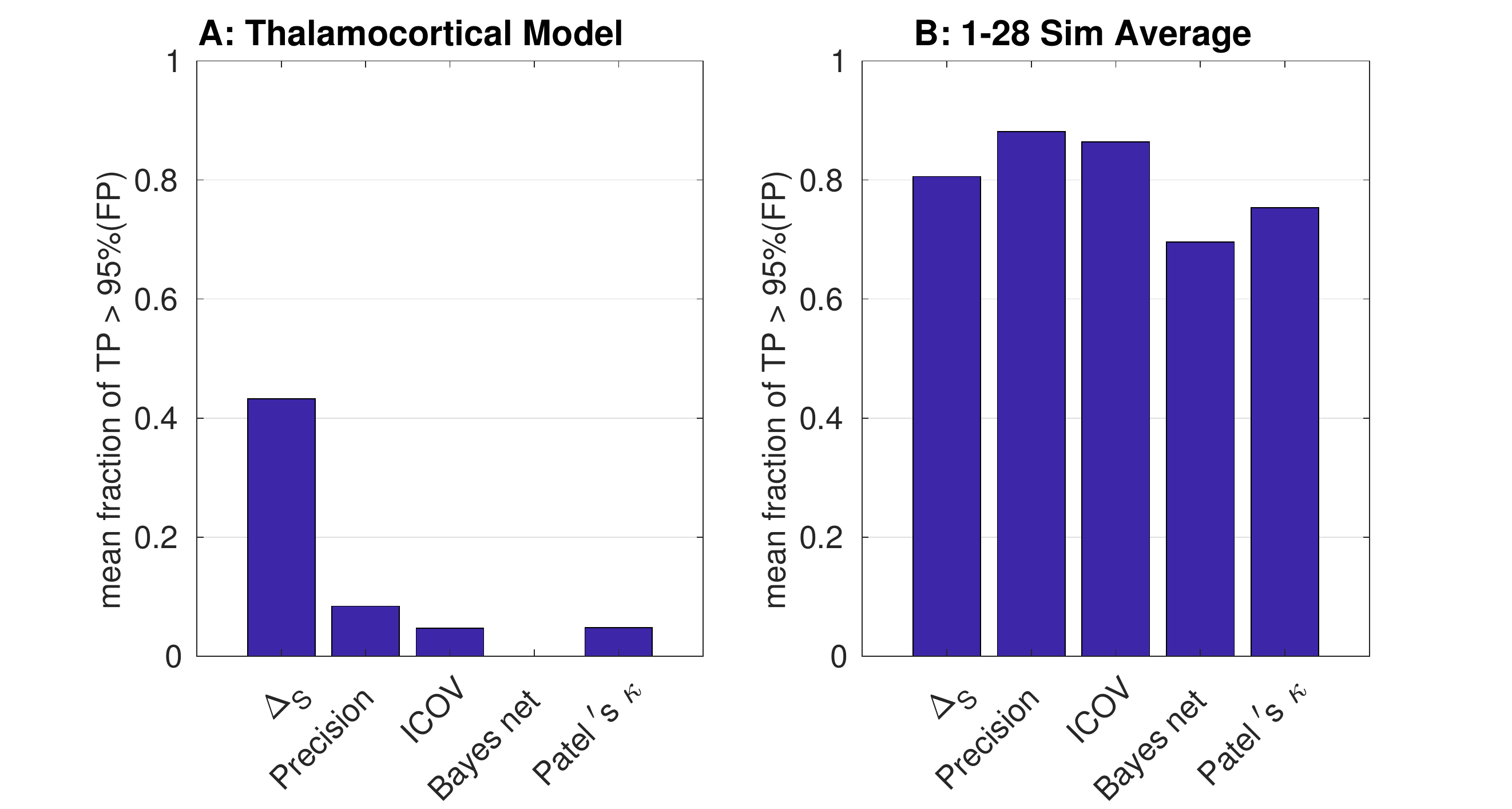}
\end{center}
\caption{Comparison of A) Performances (c-sensitivity) on our more realistic thalamocortical ionic model. B) Average performances across the 28 data sets from a previous study\citep{smith2011network}. Method's performance for each of the 28 data sets is provided in appendix~\ref{benchmarkApp}. }
\label{benchmark_ave}
\end{figure}
To show that our new method is generalizable outside of the two example simulations just given, we benchmarked various methods using a previously generated dataset containing 28 BOLD signal simulations \citep{smith2011network}. 
Top performers mentioned in the previous study \citep{smith2011network} (ICOV, Precision matrix, Bayes net, and Patel's $\kappa$) are included as benchmarks. 
Shown in Fig.~\ref{benchmark_ave}A, For the thalamocortical ionic model, we see that the $\Delta_S$ outperforms others. This is mainly because the method's better performance at reducing the type 1 false connections.

For the 28 simulations from the previous study \citep{smith2011network}, our differential covariance method's average performance is comparable to other top methods identified previously (Fig.~\ref{benchmark_ave}B). Because most of these 28 simulations only have 5 nodes and the connection matrix is not sparse, we dropped the sparse+latent regularization step in our method, except for simulation No.4, the only one that has a sparse connection matrix (50 nodes, about 70 connections).


%
%

\section{Discussion}
\label{sec_discussion}

\subsection{Main results}
The main contribution of this paper is: first, we developed a way to reconstruct the neural signals from the BOLD signals. While understanding the relationship between fMRI and neuronal activity is a well-studied topic (for reviews, see~\cite{heeger2002does,logothetis2002neural}), from Fig.~\ref{control1}, we see that this custom-designed backward model is necessary for our differential covariance method; 
second, we adapted our novel differential covariance method to estimate the functional connectivity of fMRI data.
Our main results demonstrated that our new approach for estimating the functional connectivity provides better results for the more complicated model we made. This new approach not only reduces estimation errors exist in previous methods, but also provides directionality information of the connection. 

For the DCM dataset from a previous study \citep{smith2011network}, it provides similar performance to the previously benchmarked best methods (covariance-based methods). 
As explained above, each step of the differential covariance method is specialized at reducing one type of false connections. However, since the presence of these 3 types of false connections are minimal in this dataset, our new method does not outperform the covariance-based methods. Further, because the size of the network is small, the connectivity matrix is not sparse, so we cannot take advantage of the sparse+latent regularization step in our method. We believe this will not be an issue when applying differential covariance method to real-world problems,  because the brain network has a much higher dimensionality and is sparser.

\subsection{Caveats and future directions}


As mentioned above, we made several key assumptions in this paper, and we believe these assumptions should be further tested and tuned in the future. 
Our approach depends on the neural signals reconstructed from the fMRI transfer function. Even though the fMRI transfer function we used (the Balloon model) is based on the fundamental vesicular dynamics, whether high quality neural signals can be reconstructed from the noisy real fMRI instruments should be further studied. Also the sparse and low-rank assumptions are critical to this new method. In real-world fMRI recordings, these assumptions should be validated and the scan area and sampling rate should also be tuned for best results. 

Although this paper demonstrated the validity of the differential covariance method, many improvements can be made. Compared to the covariance-based methods, this new method requires more data samples to converge, requires higher sampling rate because it needs to reconstruct the neural signals, and it has lower noise tolerance. These limitations can be alleviated in future versions of this method.

\section*{Acknowledgement}
We would like to thank Prof. Stephen Smith for generously sharing his code used in his previous work \cite{smith2011network}. We also want to thank Drs. Thomas Liu, Richard Buxton, and all members of the Computational Neurobiology Lab for providing helpful feedback. This research was supported by ONR MURI (N000141310672), Swartz Foundation and Howard Hughes Medical Institute.

\section*{Appendix}
\appendix

\section{Previous methods}
In this section we  review several top performing methods benchmarked in a previous study\citep{smith2011network}.
\subsection{covariance-based methods}
The well used covariance-based methods are used as a benchmark in this paper because a prior work \citep{smith2011network} has shown its superior performance by applying to simulated BOLD signal.
\subsubsection{covariance method}
The covariance matrix is defined as:
\begin{equation}
	COV_{x,y}=\frac{1}{N}\sum^{N}_{i=1}(x_i-\mu_{x})(y_i-\mu_{y})
\end{equation}
Where $x$ and $y$ are two variables. $\mu_{x}$ and $\mu_{y}$ are their population mean.

\subsubsection{Precision Matrix}
\label{sec_PCOV}
The precision matrix is the inverse of the covariance matrix:
\begin{equation}
P=COV^{-1},
\end{equation}
It can be considered as one kind of partial correlation. Here we briefly review this derivation, because we use it to develop our new method. The derivation here is based on and adapted from \cite{cox1996multivariate}.

We begin by considering a pair of variables $(x, y)$, and remove the correlation in them introduced from a control variable $z$. 

First, we define the covariance matrix as:
\begin{equation}
COV_{xyz}= \left[ \begin{array}{ccc}
\sigma_{xx} & \sigma_{xy}& \sigma_{xz}\\
\sigma_{yx} & \sigma_{yy}& \sigma_{yz}\\
\sigma_{zx} & \sigma_{zy}& \sigma_{zz}
\end{array} \right]
\end{equation}
By solving the linear regression problem:
\begin{equation}
\begin{array}{rcl}
w_x=\argmin\limits_{w}E(x-w*z)^2\\
w_y=\argmin\limits_{w}E(y-w*z)^2
\end{array}
\end{equation}
we have:
\begin{equation}
\begin{array}{rcl}
w_x=\sigma_{xz}\sigma_{zz}^{-1}\\
w_y=\sigma_{yz}\sigma_{zz}^{-1}
\end{array}
\end{equation}

then, we define the residual of $x,y$ as,
\begin{equation}
\begin{array}{c}
r_x = x- w_x*z\\
r_y = y - w_y*z
\end{array}
\end{equation}
Therefore, the covariance of $r_x , r_y$ is:
\begin{equation}
COV_{r_x, r_y}= \sigma_{xy} - \sigma_{xz}*\sigma_{zz}^{-1}*\sigma_{yz}
\end{equation}

On the other hand, if we define the precision matrix as:
\begin{equation}
P_{xyz}= \left[ \begin{array}{ccc}
p_{xx} & p_{xy}& p_{xz}\\
p_{yx} & p_{yy}& p_{yz}\\
p_{zx} & p_{zy}& p_{zz}
\end{array} \right]
\end{equation}

Using Cramer's rule, we have:
 \begin{equation}
p_{xy}= \frac{-\left| \begin{array}{cc}
\sigma_{xy} & \sigma_{xz}\\
\sigma_{zy} & \sigma_{zz}
\end{array} \right|}{|COV_{xyz}|}
\end{equation}
Therefore,
\begin{equation}
p_{xy}= \frac{-\sigma_{zz}}{{|COV_{xyz}|}}(\sigma_{xy} - \sigma_{xz}*\sigma_{zz}^{-1}*\sigma_{yz})
\end{equation}

So $p_{xy}$ and $COV_{r_x,r_y}$ are differed by a ratio of $ \frac{-\sigma_{zz}}{{|COV_{xyz}|}}$.

\subsubsection{Sparsely regularized precision matrix}
If the precision matrix is expected to be sparse, one way to efficiently and accurately estimate this matrix, especially when the dimension is large, is regularize the estimation using the Lasso method \citep{hsieh2014quic, hsieh2013big, friedman2008sparse, banerjee2006convex}.
 \begin{equation}
\argmin\limits_{S \succ 0} \{-logdet(S)+tr(SC) + \lambda*||S||_1\}
\end{equation}
where C is the sample covariance matrix.

We used the implementation (https://www.cs.ubc.ca/~schmidtm/Software/L1precision.html) mentioned in the  previous work \citep{smith2011network} to benchmark our new method.

\subsection{Bayes-net method}
A variety of the Bayes Net estimation methods are implemented in the Tetrad toolbox (http://www.phil.cmu.edu/tetrad/). Following the convention of the prior work \citep{smith2011network}, we tested CCD, CPC, FCI, PC and GES. 
PC ("Peter and Clark") is a causal inference algorithm used to estimate the causal relationship in a directed acyclic graph (DAG) \citep{meek1995causal}. CPC (Conservative PC) was based on the PC algorithm, but produce less false connections \citep{ramsey2012adjacency}. GES (Greedy Equivalence Search) is another implementation based on the PC algorithm, but it uses a score system to estimate the causal connections \citep{chickering2003optimal, ramsey2010six}. The FCI (Fast Causal Inference) algorithm is a different approach, which takes into account the presence of latent confounder and selection bias \citep{zhang2008completeness}. CCD (Cyclic Causal Discovery) unlike other algorithms mentioned, allows cycles in the network for estimation \citep{richardson1996automated}. The code used in this paper is based on the Tetrad Toolbox and further adapted by the authors of \cite{smith2011network} for fMRI data.

\subsection{Patel's $\kappa$ method}
The Patel's $\kappa$ measurement assess the relationship between pairs of distinct brain regions by comparing expected joint and marginal probabilities of elevated activity of voxel pairs through a Bayesian paradigm \citep{patel2006bayesian}. For this paper, we used the implementation generously provided by  the authors of \cite{smith2011network};

\subsection{Sparse latent regularization}
\label{SLreg}

Prior studies\citep{banerjee2006convex, friedman2008sparse} have shown that regularizations can provide better estimation if the ground truth connection matrix has a known structure (e.g. sparse). For all data tested in this paper, the sparse latent regularization\citep{yatsenko2015improved} worked best. For a fair comparison, we applied the sparse latent regularization to both the precision matrix method and our differential covariance method.

In the original sparse latent regularization method, people made the assumption that a larger precision matrix $S$ is the joint distribution of the $p$ observed neurons and $d$ latent neurons \citep{yatsenko2015improved}. i.e.

\[S= \left( \begin{array}{cc}
S_{11} & S_{12}\\
S_{21} & S_{22}\end{array} \right)\] 

Where $S_{11}$ corresponds to the observable neurons.
If we can only measure the observable neurons, the partial correlation computed from the observed neural signals is,
\begin{equation}
C_{ob}^{-1}=S_{ob} = S_{11}  -  S_{12}\cdot S_{22}^{-1}\cdot S_{21}
\end{equation}
because the invisible latent neurons as shown in Eq.~\ref{passiveModel} introduce correlations into the measurable system.
We denote this correlation introduced from the latent inputs as
\begin{equation}
L= S_{12}\cdot S_{22}^{-1}\cdot S_{21}
\end{equation}
If we can make the assumption that the connection between the visible neurons are sparse, i.e. $S_{11}$ is sparse and the number of latent neurons is much smaller than the number of visible neurons, i.e. $d << p$. Then, prior works \citep{chandrasekaran2011rank} have shown that if $S_{ob}$ is known, $S_{11}$ is sparse enough and L's rank is low enough (within the bound defined in \cite{chandrasekaran2011rank}), then the solution of 
\begin{equation}
S_{11}-L=S_{ob}
\end{equation}
 is uniquely defined and can be solved by the following convex optimization problem
\begin{equation}
\argmin\limits_{S_{11},L} ||S_{11}||_1 + \alpha*tr(L)
\end{equation}
under the constraint that 
\begin{equation}
S_{ob}=S_{11}-L
\end{equation}
Here, $||\ ||_1$ is the L1-norm of a matrix, and $tr()$ is the trace of a matrix. $\alpha$ is the penalty ratio between the L1-norm of $S_{11}$ and the trace of L and is set to $1/\sqrt{N}$ for all our estimations.

However, the above method is used to regularize precision matrix. For our  differential covariance estimation, we need to make small changes to the derivation. Note that
if we assume the neural signals of the latent neurons are known, and let $l$ be the indexes of these latent neurons,
then from our previous   section (section~\ref{partial_diffCov}),
\begin{equation}
\Delta_{S_{i,j}} = \Delta_{P_{i,j}} - COV_{j,l}\cdot COV_{l,l}^{-1}\cdot \Delta_{C_{i,l}}^T
\end{equation}
removes the $V_{latent}$ terms. 

Even if $l$ is unknown, 
$$COV_{j,l}\cdot COV_{l,l}^{-1}\cdot \Delta_{C_{i,l}}^T$$
is low-rank, because it is bounded by the dimensionality of $COV_{l,l}$, which is $d$. And $\Delta_S$ is the internal connections between the visible neurons, which should be a sparse matrix.
Therefore, letting 
\begin{equation}
\begin{array}{c}
S_{ob}=\Delta_P \\
S_{11}=\Delta_S \\
L= - COV_{j,l}\cdot COV_{l,l}^{-1}\cdot \Delta_{C_{i,l}}^T
\end{array}
\end{equation}
we can use the original sparse+latent method to solve for $\Delta_S$. In this paper, we used the inexact robust PCA algorithm ( \url{http://perception.csl.illinois.edu/matrix-rank/sample_code.html } ) to solve this problem \citep{lin2011linearized}.

\section{Balloon Model Parameters}
\label{balloonApp}
\begin{table}[h]
\caption{Balloon Model Parameters
}
\label{tab_1}
\resizebox{\columnwidth}{!}{
\begin{tabular}{c|c|c|c}
  Parameter  & Description & Prior Mean & Prior Variance \\ \hline  \hline
  $\kappa$ & Rate of signal decay & 0.65 per s & 0.015 \\ \hline
$\gamma$ & Rate of flow-dependent elimination & 0.41 per s & 0.002 \\ \hline
$\tau$ & Hemodynamic transit time & 0.98 per s & 0.0568 \\ \hline
$\alpha$ & Grubb’s exponent & 0.32 & 0.0015 \\ \hline
$\rho$ & Resting oxygen extraction fraction	 & 0.34 & 0.0024 \\ \hline
\end{tabular}
}
\end{table}

\section{Backward Model Parameters}
\label{backwardApp}

\begin{table}[h]
\caption{Linearized Balloon Model Parameters
}
\label{tab_2}
\resizebox{\columnwidth}{!}{
\begin{tabular}{c|c|c|c}
  Parameter  & Description & Prior Mean & Prior Variance \\ \hline  \hline
  $\kappa$ & Rate of signal decay & 0.65 per s & 0.015 \\ \hline
$\gamma$ & Rate of flow-dependent elimination & 0.41 per s & 0.002 \\ \hline
$\tau$ & Hemodynamic transit time & 0.98 per s & 0.0568 \\ \hline
$\alpha$ & Grubb’s exponent & 0.32 & 0.0015 \\ \hline
$\rho$ & Resting oxygen extraction fraction	 & 0.34 & 0.0024 \\ \hline
$\epsilon$ & Neuronal efficacy	 & 0.54 & 0.085 \\ \hline
$\tau_s$ & Signal decay	 & 1.54 & 0.169 \\ \hline
$\tau_f$ & Autoregulation	 & 2.46 & 0.212 \\ \hline
\end{tabular}
}
\end{table}

\begin{equation}
\begin{array}{lc}
k_1=7\rho\\
k_2=2\\
k_3=2\rho - 2\\
q_1=-(k_1+k_2)*\tau*\epsilon - (k_3-k_2)*\epsilon\\
q_0=-\frac{(k_1+k_2)*(\tau-1+\alpha)*\epsilon}{\alpha*\tau} - \frac{(k_3-k_2)*\epsilon}{\tau}\\
p_4=\tau\\
p_3=1+\frac{1}{\alpha} + \frac{\tau}{\tau_s}\\
p_2=\frac{1}{\tau*\alpha} + \frac{\tau}{\tau_f} + \frac{1+1/\alpha}{\tau_s}\\
p_1=\frac{1+1/\alpha}{\tau_f} + \frac{1}{\tau*\alpha*\tau_s}\\
p_0=\frac{1}{\tau*\alpha*\tau_f} 
\end{array}
\end{equation}

\section{Different method's performance for each simulation}
\label{benchmarkApp}
\begin{figure}[hbtp]
\begin{center}
\includegraphics[width=\textwidth]{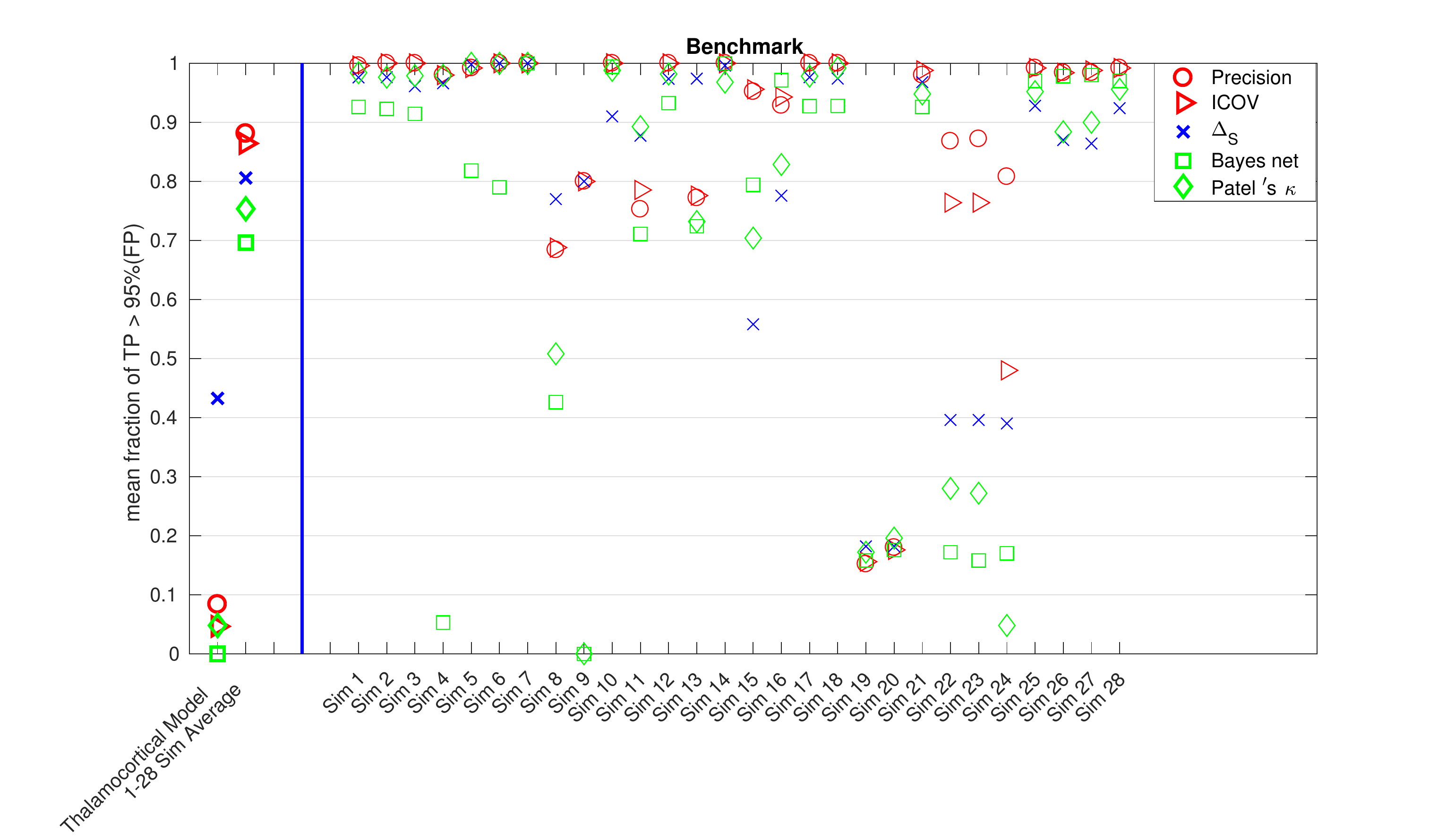}
\end{center}
\caption{The first column is method's performance (c-sensitivity) on our more realistic thalamocortical ionic model.The second column is method's average performance across the 28 data sets from a previous study\citep{smith2011network}. Column 3-30 are method's performances for each of the 28 data sets. }
\label{benchmark}
\end{figure}

\pagebreak
\bibliography{ref}
\bibliographystyle{plainnat}
\end{document}